\pdfoutput=1

\documentclass[aps,twocolumn,superscriptaddress,floatfix,preprintnumbers,nofootinbib,eqsecnum]{revtex4-1}

\usepackage{amsmath, float, graphicx,amssymb,multirow,pgfplots,pgfplotstable}
\usepackage{tikz,hyperref}
\usetikzlibrary{shapes.geometric,arrows}

\begin{document}
\title{
Explaining the ANITA Anomaly with Inelastic Boosted Dark Matter
}

 \author{Lucien Heurtier}
 \email{heurtier@email.arizona.edu}
\affiliation{Department of Physics, University of Arizona, Tucson, AZ  85721, USA}
  \author{Doojin Kim}
\email{doojinkim@email.arizona.edu}
\affiliation{Department of Physics, University of Arizona, Tucson, AZ 85721, USA}
\author{Jong-Chul Park}
\email{jcpark@cnu.ac.kr}
\affiliation{Department of Physics, Chungnam National University, Daejeon 34134, Republic of Korea}
\author{Seodong Shin}
\email{seodongshin@yonsei.ac.kr}
\affiliation{Department of Physics \& IPAP, Yonsei University, Seoul 03722, Republic of Korea}

\begin{abstract}
We propose a new physics scenario in which the decay of a very heavy dark-matter candidate which does not interact with the neutrino sector could explain the two anomalous events recently reported by ANITA.
The model is composed of two components of dark matter, an unstable dark-sector state, and a massive dark gauge boson.
We assume that the heavier dark-matter particle of EeV-range mass is distributed over the galactic halo and disintegrates into a pair of lighter -- highly boosted -- dark-matter states in the present universe which reach and penetrate the Earth. The latter scatters {\it in}elastically off a nucleon and produces a heavier dark-sector unstable state which subsequently decays back to the lighter dark matter along with hadrons, which induce Extensive Air Showers, via on-/off-shell dark gauge boson.
Depending on the mass hierarchy within the dark sector, either the dark gauge boson or the unstable dark-sector particle can be long-lived, hence transmitted significantly through the Earth.
We study the angular distribution of the signal and show that our model favors emergence angles in the range $\sim 25^\circ -35^\circ$ if the associated parameter choices bear the situation where the mean free path of the boosted incident particle is much larger than the Earth diameter while its long-lived decay product has a decay length of dimensions comparable to the Earth radius.
Our model, in particular, avoids any constraints from complementary neutrino searches such as IceCube or the Auger observatory.
\end{abstract}

\maketitle


\newcommand{\PRE}[1]{{#1}} 
\newcommand{\ul}{\underline}
\newcommand{\del}{\partial}
\newcommand{\nbox}{{\,\lower0.9pt\vbox{\hrule \hbox{\vrule height 0.2 cm
\hskip 0.2 cm \vrule height 0.2 cm}\hrule}\,}}

\newcommand{\postscript}[2]{\setlength{\epsfxsize}{#2\hsize}
   \centerline{\epsfbox{#1}}}
\newcommand{\gweak}{g_{\text{weak}}}
\newcommand{\mweak}{m_{\text{weak}}}
\newcommand{\mplanck}{M_{\text{Pl}}}
\newcommand{\mstar}{M_{*}}
\newcommand{\sigmaan}{\sigma_{\text{an}}}
\newcommand{\sigmatot}{\sigma_{\text{tot}}}
\newcommand{\sigmaSI}{\sigma_{\rm SI}}
\newcommand{\sigmaSD}{\sigma_{\rm SD}}
\newcommand{\OmegaM}{\Omega_{\text{M}}}
\newcommand{\OmegaDM}{\Omega_{\text{DM}}}
\newcommand{\ipb}{\text{pb}^{-1}}
\newcommand{\ifb}{\text{fb}^{-1}}
\newcommand{\iab}{\text{ab}^{-1}}
\newcommand{\ev}{\text{eV}}
\newcommand{\kev}{\text{keV}}
\newcommand{\mev}{\text{MeV}}
\newcommand{\gev}{\text{GeV}}
\newcommand{\tev}{\text{TeV}}
\newcommand{\pb}{\text{pb}}
\newcommand{\mb}{\text{mb}}
\newcommand{\cm}{\text{cm}}
\newcommand{\m}{\text{m}}
\newcommand{\km}{\text{km}}
\newcommand{\kg}{\text{kg}}
\newcommand{\g}{\text{g}}
\newcommand{\s}{\text{s}}
\newcommand{\yr}{\text{yr}}
\newcommand{\Mpc}{\text{Mpc}}
\newcommand{\etal}{{\em et al.}}
\newcommand{\eg}{{\em e.g.}}
\newcommand{\ie}{{\em i.e.}}
\newcommand{\ibid}{{\em ibid.}}
\newcommand{\Eqref}[1]{Equation~(\ref{#1})}
\newcommand{\secref}[1]{Sec.~\ref{sec:#1}}
\newcommand{\secsref}[2]{Secs.~\ref{sec:#1} and \ref{sec:#2}}
\newcommand{\Secref}[1]{Section~\ref{sec:#1}}
\newcommand{\appref}[1]{App.~\ref{sec:#1}}
\newcommand{\figref}[1]{Fig.~\ref{fig:#1}}
\newcommand{\figsref}[2]{Figs.~\ref{fig:#1} and \ref{fig:#2}}
\newcommand{\Figref}[1]{Figure~\ref{fig:#1}}
\newcommand{\tableref}[1]{Table~\ref{table:#1}}
\newcommand{\tablesref}[2]{Tables~\ref{table:#1} and \ref{table:#2}}
\newcommand{\Dsle}[1]{\slash\hskip -0.28 cm #1}
\newcommand{\met}{{\Dsle E_T}}
\newcommand{\mpt}{\not{\! p_T}}
\newcommand{\Dslp}[1]{\slash\hskip -0.23 cm #1}
\newcommand{\Dsl}[1]{\slash\hskip -0.20 cm #1}

\newcommand{\mB}{m_{B^1}}
\newcommand{\mq}{m_{q^1}}
\newcommand{\mf}{m_{f^1}}
\newcommand{\mKK}{m_{KK}}
\newcommand{\WIMP}{\text{WIMP}}
\newcommand{\SWIMP}{\text{SWIMP}}
\newcommand{\NLSP}{\text{NLSP}}
\newcommand{\LSP}{\text{LSP}}
\newcommand{\mWIMP}{m_{\WIMP}}
\newcommand{\mSWIMP}{m_{\SWIMP}}
\newcommand{\mNLSP}{m_{\NLSP}}
\newcommand{\mchi}{m_{\chi}}
\newcommand{\mgravitino}{m_{\gravitino}}
\newcommand{\mmed}{M_{\text{med}}}
\newcommand{\gravitino}{\tilde{G}}
\newcommand{\Bino}{\tilde{B}}
\newcommand{\photino}{\tilde{\gamma}}
\newcommand{\stau}{\tilde{\tau}}
\newcommand{\slepton}{\tilde{l}}
\newcommand{\snu}{\tilde{\nu}}
\newcommand{\squark}{\tilde{q}}
\newcommand{\mgaugino}{M_{1/2}}
\newcommand{\epsEM}{\varepsilon_{\text{EM}}}
\newcommand{\mmess}{M_{\text{mess}}}
\newcommand{\lmess}{\Lambda}
\newcommand{\nmess}{N_{\text{m}}}
\newcommand{\signmu}{\text{sign}(\mu)}
\newcommand{\Omegachi}{\Omega_{\chi}}
\newcommand{\lambdafs}{\lambda_{\text{FS}}}
\newcommand{\be}{\begin{equation}}
\newcommand{\ee}{\end{equation}}
\newcommand{\bea}{\begin{eqnarray}}
\newcommand{\eea}{\end{eqnarray}}
\newcommand{\beq}{\begin{equation}}
\newcommand{\eeq}{\end{equation}}
\newcommand{\beqn}{\begin{eqnarray}}
\newcommand{\eeqn}{\end{eqnarray}}
\newcommand{\baln}{\begin{align}}
\newcommand{\ealn}{\end{align}}
\newcommand{\lsim}{\lower.7ex\hbox{$\;\stackrel{\textstyle<}{\sim}\;$}}
\newcommand{\gsim}{\lower.7ex\hbox{$\;\stackrel{\textstyle>}{\sim}\;$}}

\newcommand{\ssection}[1]{{\em #1.\ }}
\newcommand{\rem}[1]{\textbf{#1}}

\def\ie{{\it i.e.}\/}
\def\eg{{\it e.g.}\/}
\def\etc{{\it etc}.\/}
\def\calN{{\cal N}}

\def\mptwo{{m_{\pi^0}^2}}
\def\mp{{m_{\pi^0}}}
\def\sqtsn{\sqrt{s_n}}
\def\sqtsn{\sqrt{s_n}}
\def\sqtsn{\sqrt{s_n}}
\def\sqts0{\sqrt{s_0}}
\def\Dsqts{\Delta(\sqrt{s})}
\def\Omegatot{\Omega_{\mathrm{tot}}}


\section{Introduction}

The ANtarctic Impulsive Transient Antenna (ANITA) Collaboration recently reported two anomalous upward-moving cosmic-ray-like events whose inferred energy lies in $\sim 0.5 - 1$ EeV~\cite{Gorham:2016zah,Gorham:2018ydl}.
Within the framework of the Standard Model (SM), one could envision, for example, an EeV-scale tau-neutrino traversing the Earth and converting to a tau lepton which is responsible for producing an Extensive Air Showers (EAS) in the atmosphere after escaping the surface of the Earth.
However, given the typical amount of energy that such a neutrino would have to carry, and the interaction strength of the latter with nuclei, its propagation through the Earth on large distances is extremely unlikely to let a subsequent tau lepton escape the Earth at a sizable emergence angle (see also FIG.~\ref{fig:coord} for notational conventions).
Indeed, a boosted tau neutrino having an energy of $\mathcal O(1)~\mathrm{EeV}$ has a transmission probability lower than $10^{-6}$ to induce an EeV-scale tau lepton even if one includes multiple $\tau -\nu_\tau$ regenerations~\cite{Gorham:2018ydl}, and a dedicated study in Ref.~\cite{Fox:2018syq} has concluded that SM explanations are excluded by (at least) $5\sigma$ confidence.
Furthermore, Ref.~\cite{Romero-Wolf:2018zxt} explored the hypothesis of a  tau-neutrino origin together with a dedicated acceptance estimate, and reached the conclusion that the ANITA events disfavor a diffuse tau-neutrino interpretation in comparison with the Auger~\cite{Aab:2015kma} and IceCube~\cite{Aartsen:2016ngq} upper limits.
This suggests that new physics should come into play in order to accommodate the anomaly in a self-consistent manner, stimulating the theory community to propose various beyond-the-SM (BSM) scenarios.

For completeness, we should mention that the ANITA anomalous events might be explained in pure SM interpretations which were recently proposed.
A downward-moving high-energy cosmic rays might indeed be able to produce the ANITA's inverted polarity signals through coherent transition radiation from the geomagnetic-induced current in the associated air showers~\cite{deVries:2019gzs} or could be reflected by sub-layer of the ice-sheet of the Antartic \cite{Shoemaker:2019xlt}.
Such proposals remain for now speculative and still require experimental confirmations, in particular from the ANITA Collaboration itself, and therefore we shall focus in this work on the possibility that such events are a sign of new physics.

Among other attempts, several trials assumed the existence of non-SM neutrino species in order to address the signal. For example, the authors of Ref.~\cite{Cherry:2018rxj,Huang:2018als} adopt sterile neutrinos induced by Ultra High Energy (UHE) cosmic rays as ``agents'' propagating through the Earth in order to explain the ANITA events, while Ref.~\cite{Chauhan:2018lnq} interprets the signal with lepto-quark-mediated sterile neutrinos.
By contrast, Ref.~\cite{Anchordoqui:2018ucj} hypothesizes super-heavy right-handed neutrinos captured in the Earth which subsequently decay to active neutrinos in the vicinity of the Earth surface.
Some of the latest attempts imagine new physics scenarios within the supersymmetric framework.
Examples include interpretations utilizing next-to-lightest supersymmetric tau lepton (i.e, ``stau'')~\cite{Fox:2018syq}, long-lived bino in R-parity violating supersymmetry~\cite{Collins:2018jpg}, and neutrino-initiated supersymmetric sphaleron transitions~\cite{Anchordoqui:2018ssd}.
In contrast to the approaches taken in preceeding works, the authors of Refs.~\cite{Heurtier:2019git} and \cite{Hooper:2019ytr} explored the possibility that the flux of boosted particles necessary to explain the signal is due to the decay of a super-heavy dark-matter particle in the Milky Way. In \cite{Heurtier:2019git} the ANITA anoumalous events are explained with a dark-matter species decaying into a pair  of right-handed neutrinos, which  later on convert into taus inside the Earth and produce high-energy EAS in the atmosphere. In a complementary manner, also using the decay of a super-heavy dark-matter particle, the authors of Ref.~\cite{Hooper:2019ytr} proposed an interpretation in terms of the Askaryan emission created by elastic scattering of feebly interacting particles within the ice sheet. The authors of Ref.~\cite{Cline:2019snp} finally reviewed different topologies of processes which could explain the signal with EAS and proposed a model in which a heavy dark-matter decays into a boosted light component which converts into taus after scattering within the Earth.
As another recent BSM interpretation, the authors of Ref.~\cite{Esteban:2019hcm} argued that the reflection of radio pulses could reproduce the ANITA signals via axion-photon conversion.

As a generic feature of most of the aforementioned interpretations, we point out that previous works rely on the interaction which might exist between new-physics states and ordinary SM neutrinos in order to induce the relevant experimental signatures. The new states thus {\it effectively} ``transport'' the neutrinos through the Earth to enhance the chance that they convert into leptons in the vicinity of the surface.

While producing leptons inside the Earth from the conversion of neutrinos is probably the most natural SM-{\it friendly} interpretation at low energy, the situation is radically different at very high energy since the neutrino is not a good candidate to help propagating through the Earth on large distances. Moreover, there are no compelling reasons that neutrinos are necessary ingredients to explain the ANITA anomaly. Indeed, the only requirement for ANITA to detect a signal is that a significant electric field is locally produced around the detector. Therefore, as far as EAS are concerned, the production of any boosted hadronic final states in the atmosphere, whatever its parent particle is, would suffice to propose an interpretation for the anomalous events reported by the Collaboration.

Furthermore, the connection between the neutrino sector and new-physics states has been thoroughly probed by different experiments in the past and eventually leads to tensions between existing experimental constraints and the parameter choices required in order to fit the data.
In particular, the production of highly boosted -- and therefore very long-lived -- $\tau$ leptons close to the Earth surface, which can pass through terrestrial detectors such as IceCube before decaying in the atmosphere, may leave sizable signatures which should already have been observed.

In light of these considerations, we propose a dark-matter scenario not accompanying the production of SM leptons inside the Earth (in contrast to that in \cite{Heurtier:2019git} and \cite{Hooper:2019ytr}) in order to explain the two ANITA anomalous events.
The model that we consider is built upon a non-minimal dark sector scenario, containing
\begin{itemize}
\item $\phi$, a super-heavy scalar dark matter,
\item $\chi_1$, a light fermionic dark-matter component,
\item $\chi_2$, an unstable fermionic dark-sector state, and
\item $X$, a dark gauge boson
interacting with the aforementioned fermionic states and mixing with the SM photon through kinetic mixing.
\end{itemize}

\begin{figure*}[t]
\centering
\includegraphics[width=16cm]{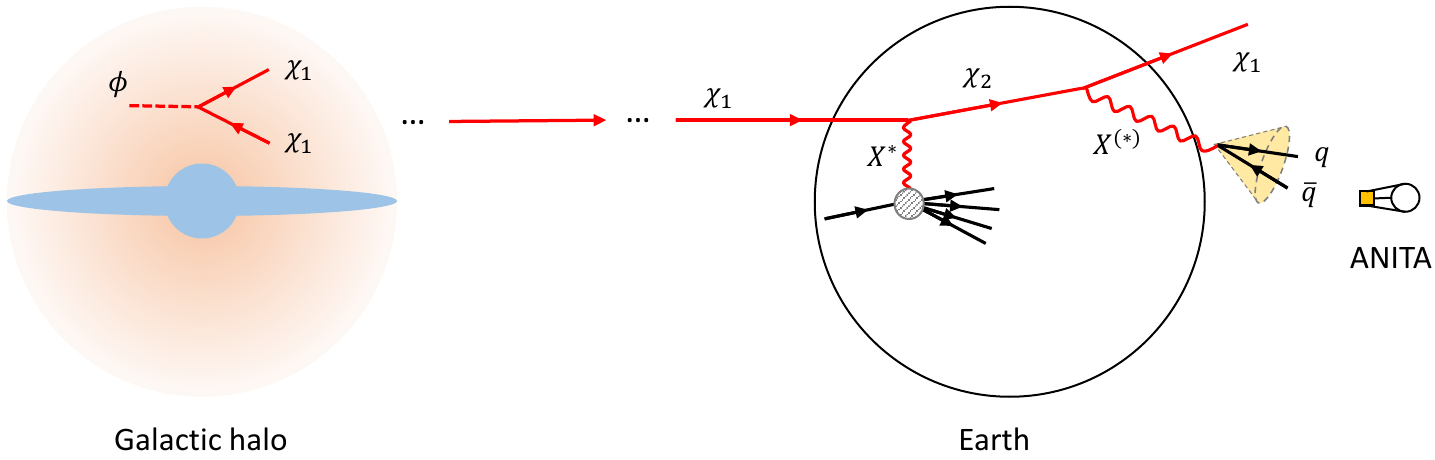}
\caption{\label{fig:signature} The benchmark dark-matter scenario under consideration for explaining the ANITA anomaly.
The travel lengths of individual particles and the angles between particles are randomly chosen for illustration; that is, they do not represent any ANITA-related events at all.}
\end{figure*}

We assume that the whole signal process sketched in FIG.~\ref{fig:signature} begins with the decay of $\phi$ into a pair of $\chi_1$'s over the galactic halo in the present universe.\footnote{In principle, $\phi \to \chi_1 + Y$ with $Y$ collectively representing all other particles is possible, but we take $\phi \to \chi_1\chi_1$ as a benchmark model.}
Each $\chi_1$ then acquires a large Lorentz boost factor due to the mass gap between the two dark matter species.\footnote{Similar scenarios in which the decay of a heavier dark matter component boosts a pair of lighter dark matter components have been discussed in the context of explaining the IceCube PeV-neutrino events~\cite{Bhattacharya:2014yha,Kopp:2015bfa}.}
Such boosted $\chi_1$ scatters off material in the Earth to the unstable dark-sector state $\chi_2$, via a $t$-channel exchange of the dark gauge boson $X$.
By construction, $\chi_2$ is heavier than $\chi_1$, and hence decays back to $\chi_1$ along with a quark pair whose hadronization eventually creates an EAS for the ANITA detector as long as the production happens in the atmosphere.
In this scenario, the particle $\chi_1$ behaves like a neutrino, as it is not only invisible but also ultra-relativistic, while the model parameters associated with $\chi_1$ are relatively less constrained by existing experiments.
Depending on the mass spectrum among $\chi_1$, $\chi_2$ and $X$, either $X$ or $\chi_2$ can be naturally long-lived so that the quark pair $q\bar{q}$ can emerge in the atmosphere with a sizable probability.
We shall discuss both of the possibilities in our analysis and demonstrate that they successfully accommodate the ANITA events, yielding the highest event probability at an emergence angle $\theta_{\rm em}$ of $25^\circ -35^\circ$ around which the two reported events are associated.

We additionally provide generic insights in order to understand the angular distribution of the events predicted for ANITA in the general case where a long-lived particle from the dark sector would be produced through scattering of a boosted dark-sector particle in the Earth and decay into hadronic final stats in the atmosphere.
In particular, we show that the situation where the mean free path of the boosted incoming particle is much larger than the Earth diameter but where its decay product has a mean decay length of dimensions comparable to the Earth radius can lead to an interesting ``translucent'' case in which the angular distribution can culminate at moderately large emergence angles.

The paper is structured as follows.
In Sec.~\ref{sec:transparency} we start by developing general arguments and demonstrate how the Earth transparency can help predicting events for ANITA with relatively large emergence angles. In Sec.~\ref{sec:model}, we define our benchmark model which may give rise to the signal of interest.
Our analysis procedure including relevant Monte Carlo simulation is detailed in Sec.~\ref{sec:analysis}.
We then report our main findings and results in Sec.~\ref{sec:results}, followed by some discussions and interpretations.
Section~\ref{sec:conclusion} is reserved for conclusions and future prospects.

\section{Understanding the Angular Distribution of EAS}\label{sec:transparency}

The ANITA Collaboration has thus far collected 85.5-day data for three different flights, and recorded more than 30 interesting events.
Of them, 2 events are identified upward-moving with an energy of $\sim 0.6$ EeV, as they do not show a phase reversal characteristic of EAS.
Hence, high-energetic upward-propagating cosmic rays may induce these events. Event specifications in terms of reconstructed energy $E_{\rm rec}$ and emergence angle $\theta_{\rm em}$ are summarized below, while we refer to \cite{Gorham:2016zah,Gorham:2018ydl} for more details.
\begin{itemize}
\item Event \#3,985,267 in ANITA I: $E_{\rm rec}=0.6\pm 0.4$ EeV and $\theta_{\rm em} =25.4\pm 1^\circ$

\item Event \#15,717,147 in ANITA III: $E_{\rm rec}=0.56^{+ 0.3}_{-0.2}$ EeV and $\theta_{\rm em} =35.5\pm 1^\circ$
\end{itemize}

One of the most puzzling features of these anomalous events, besides the propagation of very high-energy particles on large distances, is the angular distribution of such events.
While most of the existing interpretations simply rely on the total effective area of the detector in order to compute a total number of events for a certain astrophysical flux, Refs.~\cite{Cherry:2018rxj,Huang:2018als,Heurtier:2019git,Hooper:2019ytr,Cline:2019snp} have attempted to describe the angular repartition of the events predicted by an incoming flux of sterile/right-handed neutrinos.
However, the distributions in these studies generically favor relatively small emergence angles ($\lesssim 10^\circ$), even though they provide an explanation for the number of events observed by the ANITA Collaboration at angles larger than SM neutrinos would do.

In light of this phenomenological challenge, we discuss what class of scenarios would preferentially give rise to events with a relatively large emergence angle.
As detailed in Sec.~\ref{sec:model}, all the ingredients in our benchmark model, the incident particle arriving on the Earth, the particles propagating through the Earth, and the particle inducing an EAS in the low atmosphere, belong to a dark sector.
In this section we shall develop some fairly model-independent insight about how the scenario of this sort can give rise to an angular distribution of EAS which is in favor of smaller or larger emergence angles recorded by ANITA.
To this end, we simplify the situation at hand, assuming that a particle $A$ scatters off a nucleon and produces a particle $B$ which propagates through the Earth and decays into pairs of quarks:
\begin{equation}\label{eq:process}
A\stackrel{\sigma_{AN}}{\xrightarrow{\hspace*{1.cm}}} B \stackrel{\Gamma_B}{\xrightarrow{\hspace*{1.cm}}} \bar q q\,,
\end{equation}
where $\sigma_{AN}$ and $\Gamma_B$ parameterize the scattering process of $A$ and the decay of $B$, respectively.
As described in more detail in Sec.~\ref{sec:analysis}, the effective area of the ANITA detector is a convolution of following different effects:
\begin{enumerate}
\item The bigger
$\sigma_{AN}$ is, the more $B$ particles are produced inside the Earth for a given incoming flux of $A$.
Thereafter, the longer-lived the $B$ particle is, the more probable it decays after escaping the surface rather than inside the Earth.
\item Once escaping the Earth, the shorter-lived the $B$ particle is, the more it decays before passing the ANITA detector.
\item Depending on the location of the decay and the opening angle between the ANITA detector and the shower axis, the resulting electric field may initiate detection of an event.
\end{enumerate}

The effective area of the detector per unit emergence angle can then be sketched to be of the form
\begin{eqnarray}
dA_{\rm eff}&\approx& d\Omega_E(\vec{r}_{\rm ch})\otimes P_{\rm ex}(\theta_{\rm em},\mathcal{E}_{\rm ex})\nonumber\\
&&\otimes P_{\rm dec}(\theta_{\rm em},\mathcal{E}_{\rm ex}) \otimes P_{\rm det}(\vec{r}_{\rm ch})\,,
\end{eqnarray}
where $\vec{r}_{\rm ch}$ denotes the spatial direction of the chord of interest and where $d\Omega_E$ corresponds to the Earth elementary area per unit  emergence angle (integrated over the azimuthal angle around the rotation axis of the Earth).
The quantity $P_{\rm ex}(\theta_{\rm em},\mathcal{E}_{\rm ex})$ denotes the probability that a particle $B$ escapes the Earth with an energy of $\mathcal{E}_{\rm ex}$ for a given incoming particle $A$ (step 1.).
Finally, $P_{\rm dec}(\theta_{\rm em},\mathcal{E}_{\rm ex})$ and $P_{\rm det}(\vec{r}_{\rm ch})$ respectively stand for the probability that the particle $B$ decays in the atmosphere (step 2.) and the probability that the shower produced by such a decay is detected by ANITA (step 3.).

\begin{figure}[t]
\begin{center}
\includegraphics[width=0.9\linewidth]{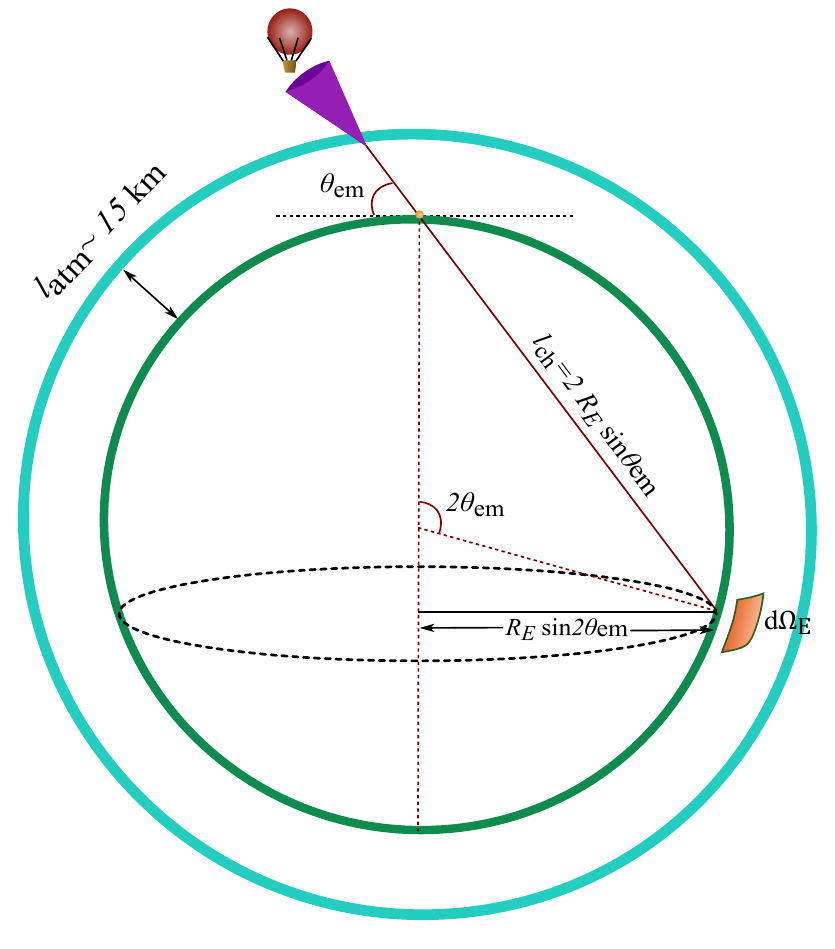}
\caption{\label{fig:geometry}Geometry of the chord propagation, as a function of the emergence angle $\theta_{\rm em}$.
The chord propagation and the Earth radius join at an angle of $\pi/2 -\theta_{\rm em}$.}
\end{center}
\end{figure}

For simplicity, we assume for now that the detected showers are so boosted that we can consider only the chords crossing the ANITA location and simply look at the variation of these quantities as a function of the emergence angle only.
In our scenario, the particles $A$ and $B$
are so feebly interacting that we can neglect any loss of energy which could arise while propagating.
Therefore, the exit energy $\mathcal{E}_{\rm ex}$ can be uniquely related to the energy of the incoming $A$ particle.

The elementary solid angle contributing to the effective area, integrated over the azimuthal angle, for a given emergence angle (see FIG.~\ref{fig:geometry}) is
\begin{equation}\label{eq:solidangle}
R_E^2 d\Omega_E(\vec{r}_{\rm ch})=4\pi R_E^2 \sin \theta_{\rm em} \sin 2\theta_{\rm em} d\theta_{\rm em}\,,
\end{equation}
where $R_E$ stands for the radius of the Earth.
Therefore, neglecting the altitude of the ANITA radio-antenna, we see that a particle -- which would ideally propagate freely through the Earth, decay in the low atmosphere with probability being one, and produce a sufficiently energetic EAS -- would be detected by ANITA with a total effective area equal roughly to the Earth area, which is a strength of the ANITA detector.
Obviously, Eq.~\eqref{eq:solidangle} implies that the angular distribution of the events detected by ANITA would be necessarily suppressed around $\theta_{\rm em}=0^\circ$ and $\theta_{\rm em}=90^\circ$ while showing a peak at $\theta_{\rm em} =54.7^\circ$.

However, in the presence of interactions in the Earth, the interaction probability of the particle $A$ along a chord of length $l_\text{ch}$ varies as
\be\label{eq:interactionproba}
\frac{d P_\text{scat.}}{d l}(l<l_\text{ch})=\frac{1}{l_A}\exp\left(-\frac{l}{l_A}\right) \,,
\ee
where $l_A$ denotes the mean free path of $A$.
This distribution sharply peaks around $l\ll l_{ch}$ in the case of $l_\text{ch}\gg l_A$, but becomes relatively flat in the case of $l_\text{ch}\ll l_A$.
Therefore, the shorter $l_A$ is, the more the scattering processes happen immediately after $A$ enters the Earth surface.

In the former case (i.e., $l_\text{ch}\gg l_A$), the efficiency of the scattering process would then be almost independent of the emergence angle.
Thereafter, if the particle $B$ produced by such a scattering is long-lived enough, it would escape the Earth with a probability being higher at low emergence angles, since the probability that it decays inside the Earth is smaller for small propagation lengths as compared to particles produced deep under the surface.
Such a scenario would therefore be disfavored by the fact that ANITA has seen anomalous events at relatively large emergence angles.

\begin{figure}[t]
\begin{center}
\includegraphics[width=\linewidth]{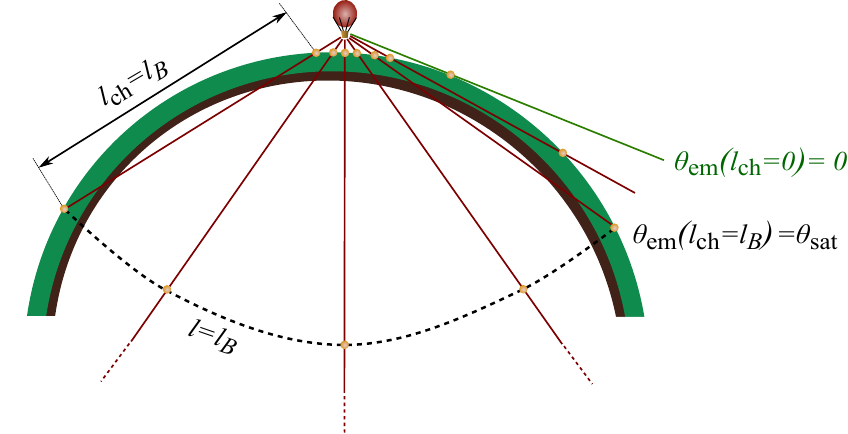}
\caption{\label{fig:chords}Region of the Earth crust inside which the scattering of a feebly interacting particle $A$ produces a particle $B$ of decay length $l_B$ which would have a significant probability to escape the Earth if propagating on a chord of length $l_{\rm ch}\lesssim l_B$.}
\end{center}
\end{figure}

\begin{figure}[t]
\begin{center}
\includegraphics[width=\linewidth]{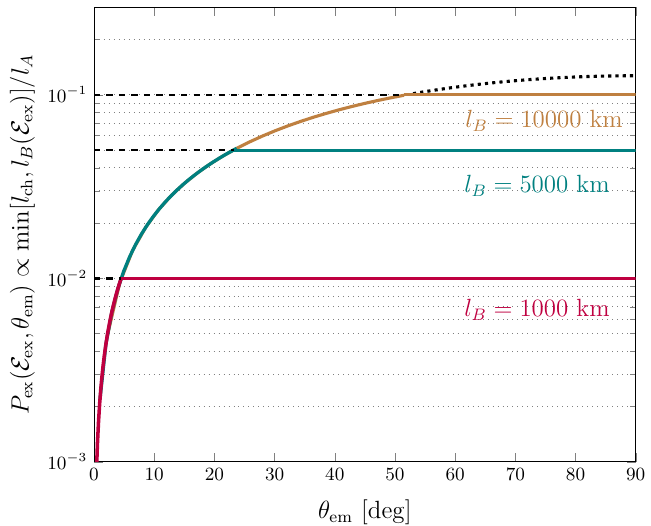}
\caption{\label{fig:Lchord}Sketch of the shape of the exit probability function $P_{\rm ex}$ for $l_B \lesssim l_{\rm ch} \ll l_A$, according to the piecewise function given in Eq.~\eqref{eq:probaTransLucent}.
$l_A$ is fixed to $10^5$ km for illustration.
Shown are the cases with three representative values of the decay length $l_B$, one smaller than, one comparable to, and one larger than the Earth radius by purple, teal, and brown curves, respectively.   }
\end{center}
\end{figure}

By contrast, in the latter case (i.e., $l_\text{ch}\ll l_A$), the Earth is almost transparent to the incoming particles so that the probability in Eq.~\eqref{eq:interactionproba} can be approximated to
\be\label{eq:interactionflat}
\frac{d P_\text{scat.}}{d l_\text{ch}}(l<l_\text{ch})\approx \frac{1}{l_A}\,.
\ee
This implies that the interaction probability along a chord of $l_\text{ch}$ is simply $l_\text{ch}/l_A$.
If particle $B$ is long-lived enough, then the exit probability becomes (roughly) proportional to $P_\text{scat.}$, i.e.,
\be
P_{\rm ex}(\theta_{\rm em},\mathcal{E}_{\rm ex})\propto \frac{l_\text{ch}}{l_A} \quad \hbox{for }l_A\gg l_\text{ch}\,,
\ee
and the relation is valid as far as the decay length of $B$ is larger than the chord length.
This shows that the resulting probability distribution is a linearly increasing function in $l_{\rm ch}$ or equivalently in $\theta_{\rm em}$.\footnote{Note that the effect is accentuated as the mean density along the chord is larger when the chord length increases.}
On the other hand, if particle $B$ is relatively short-lived (i.e., its decay length is shorter than the chord length), the event where $A$ scatters at a depth larger than the $B$ decay length would not allow $B$ to escape the Earth surface since it is likely to decay earlier.
The volume of the Earth contributing efficiently to the production of $B$ which may escape is therefore delimited by the surface $l=\min(l_{\rm ch},l_B)$ as depicted in FIG.~\ref{fig:chords}.
The exit probability is then scaling like
\be\label{eq:probaTransLucent}
P_{\rm ex}(\theta_{\rm em},\mathcal{E}_{\rm ex})\propto \frac{\min [l_\text{ch},l_B(\mathcal{E}_{\rm ex})]}{l_A} \quad \hbox{for } l_B\lesssim l_\text{ch}\,.
\ee
In other words, the proportionality of $P_{\rm ex}$ simply follows $l_{\rm ch}$ for $l_{\rm ch}<l_B$ and is frozen to $l_{\rm ch}=l_B$ for $l_{\rm ch}>l_B$.
This implies that if the decay product is short-lived as compared to the radius of the Earth, the exit probability has to saturate to a constant value, independent of the emergence angle $\theta_\text{em}$ as is schematically shown in FIG.~\ref{fig:Lchord} with illustrative values of the decay length $l_B$.

As far as the decay probability is concerned, for a  particle $B$ escaping the Earth with an emergence angle $\theta_{\rm em}$, the probability that $B$ decays before reaching an altitude $h_{\rm atm}\sim 15~\mathrm{km}$ is given by
\begin{equation}
P_{\rm dec}(\theta_{\rm em},\mathcal{E}_{\rm ex})=1-\exp\left(-\frac{l_{\rm atm}(\theta_{\rm em})}{l_B(\mathcal{E}_{\rm ex})}\right)\,,
\end{equation}
where
\begin{eqnarray}
l_{\rm atm}(\theta_{\rm em})&=& -R_E \sin\theta_{\rm em}\nonumber\\
&+& \sqrt{(R_E + h_{\rm atm})^2 - R_E^2 \cos^2\theta_{\rm em}}\,.
\end{eqnarray}

\begin{figure}[t]
\begin{center}
\includegraphics[width=0.6\linewidth]{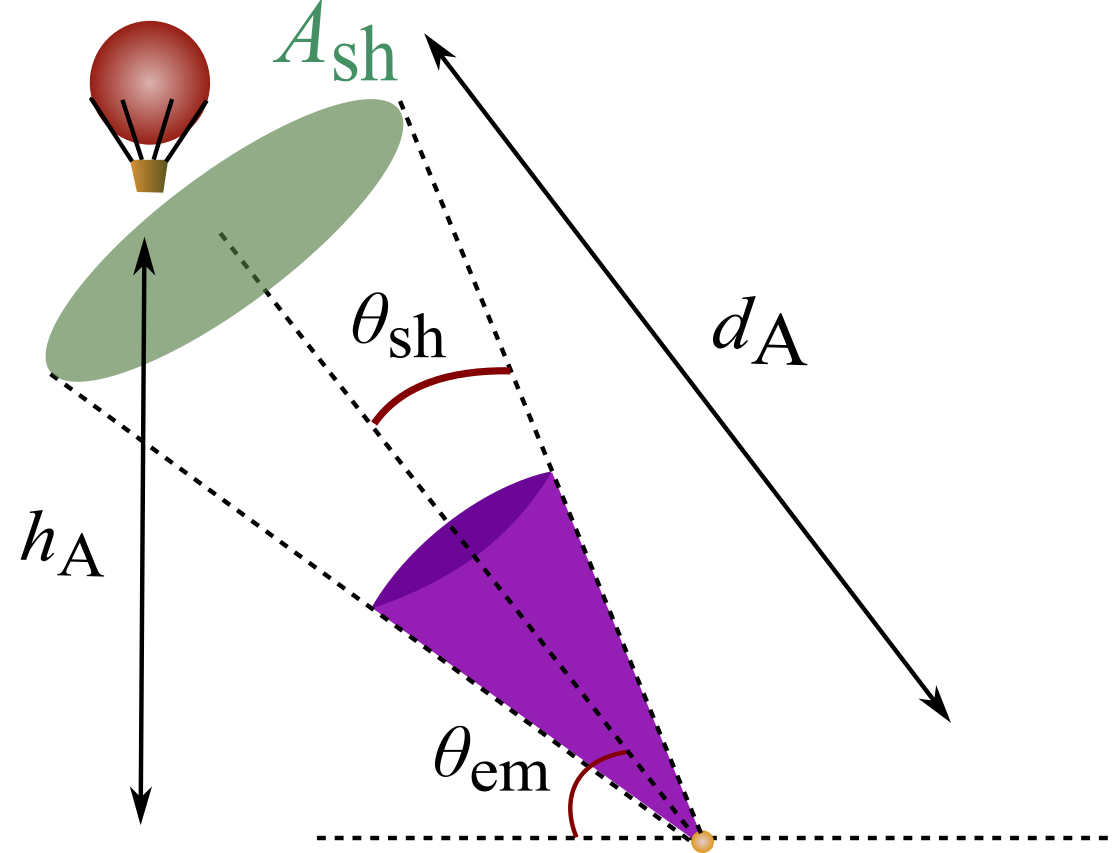}
\caption{\label{fig:opening} Notational conventions for the geometry of the shower opening.}
\end{center}
\end{figure}

Once the shower created in the lower atmosphere, with a maximum angular opening $\theta_{\rm sh}$, we assume that in our analysis ANITA can observe the signal as long as the detector is located inside the cone originating at the point of the decay, as depicted in FIG.~\ref{fig:opening}.
Therefore, the capability of detecting a shower, encoded in $P_{\rm det}$ for a given emergence angle, need to be weighted by the surface area of the shower $A_{\rm sh}$ at the detector location which is, in turn, quadratically proportional to the distance between the ANITA detector and the decay point $d_{\rm A}$:
\begin{eqnarray}
P_{\rm det}&\propto& A_{\rm sh}\propto d_{\rm A}(\theta_{\rm em})^2 = \nonumber\\
&& \left(-R_E \sin\theta_{\rm em}
+ \sqrt{(R_E + h_{\rm A})^2 - R_E^2 \cos^2\theta_{\rm em}}\right)^2\,,\nonumber\\
\end{eqnarray}
where $h_{\rm A}(=35~\mathrm{km})$ is the altitude of the ANITA detector above the Earth. However, the reader should note that if showers are emitted too far from the detector, the resultant electric field amplitude may be diminished and as a consequence the detector may not be triggered.
A proper treatment for such an effect would require a careful simulation study on the shower electric field, which is beyond the scope of this work.
Nevertheless, we expect that it would tend to suppress even more the effective area towards lower angles in favor large emergence angles, enhancing the effect that we are describing here.

\begin{figure}[t]
\begin{center}
\includegraphics[width=\linewidth]{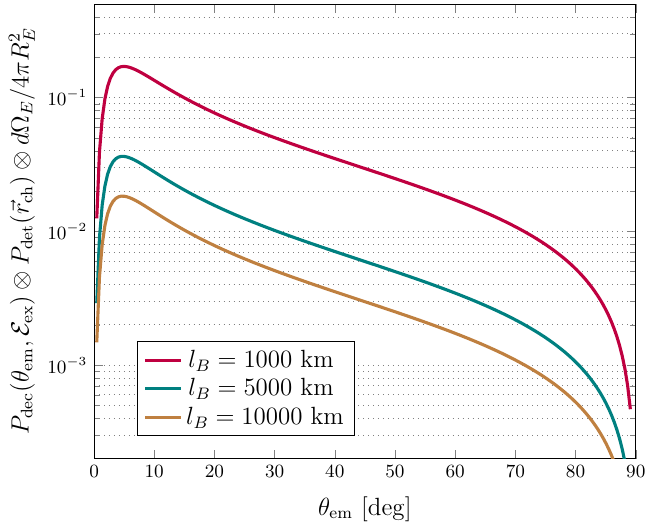}
\caption{\label{fig:decayproba} Probability that an escaping $B$ particle with emergence angle $\theta_{\rm em}$ decays within 15~km of altitude, weighted by the detection probability $P_{\rm det}(\vec{r}_{\rm ch})$ and the elementary solid angle $d\Omega_E(\vec{r}_{\rm ch})$, for the same choices of $l_B$ as in FIG.~\ref{fig:Lchord}. }
\end{center}
\end{figure}

In FIG.~\ref{fig:decayproba} we plot the differential probability, combining $P_{\rm dec}(\theta_{\rm em},\mathcal{E}_{\rm ex})$, $P_{\rm det}(\vec{r}_{\rm ch})$, and the elementary solid angle of Eq.~\eqref{eq:solidangle}.
The functional behavior of the final effective area in $\theta_{\rm em}$  is given in competition between the combined probability in FIG.~\ref{fig:decayproba} and the exit probability in FIG.~\ref{fig:Lchord}.
We now categorize various possibilities into three different regimes enumerated below.

\begin{itemize}
\item {\bf Opaque case}: In the case, similarly to the SM neutrino one, where the Earth is opaque enough to absorb most of the $B$ particles before they can escape the Earth, the exit probability is suppressed at large emergence angles (see also purple curves). Therefore, given the similar suppression of the decaying probability of $B$ exhibited in FIG.~\ref{fig:decayproba}, the effective area of ANITA will peak at a very small emergence angle.
\item {\bf Transparent case}: In the case where the Earth is transparent, both to the $A$ particle and to the decaying particle $B$, the exit probability will basically be given by the evolution of the chord length (see also brown curves). As a consequence, it will be maximized at a large angle. The suppression of the decay probability will compete only at large emergence angles, leading to an overall shape of the effective area with a maximum relatively close to the vertical (i.e., $\theta_{\rm em}=90^\circ$).
\item {\bf Translucent case}: In the intermediate case, where the Earth would be relatively transparent to $A$ but where $B$ would have a decay length comparable to the Earth radius, the shape of the effective area will possess the maximum at a value of the emergence angle around $25^\circ - 35^\circ$, which could potentially explain the value of the emergence angle of the anomalous events observed by the ANITA Collaboration (see also teal curves).
\end{itemize}
\begin{figure}[t]
\begin{center}
\includegraphics[width=\linewidth]{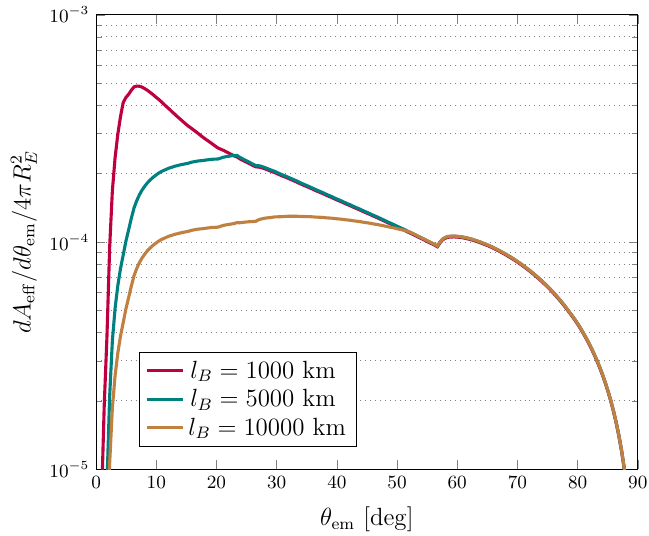}
\caption{\label{fig:Aeffsketch} Qualitative estimate for the effective area of the detector. $l_A$ is fixed to $10^5$ km and three different $l_B$ values are chosen.
For completeness, probability values are additionally weighted by the mean density along the chord, in order to incorporate the better scattering rate at large emergence angles due to the higher density of the Earth along the propagation.}
\end{center}
\end{figure}
We also visualize this qualitative summary of the three regimes in FIG.~\ref{fig:Aeffsketch}, not only incorporating the different contributions to the effective area but taking into account the fact that different emergence angles are, in reality, affected by different mean density  along the propagation chord $\langle \rho\rangle_{\rm ch}$ (we used in our simulation the Preliminary Earth Reference Model of Ref.~\cite{dziewonski1981preliminary}).
The latter factor essentially affects the exit probability $P_{\rm ex}$. We simply re-weight it by $\langle \rho\rangle_{\rm ch}/\rho_{\rm max}$ with $\rho_{\rm max}$ being the maximum density inside the Earth as we are here interested in spectral behaviors of $A_{\rm eff}$.
From now on we explore a model of inelastic boosted dark matter, and show that viable regions of parameter space can describe the above-discussed translucent case.

\section{Benchmark Model \label{sec:model}}

We are now in the position to define our benchmark dark-matter model to explain the ANITA anomaly.
Once the model definition is established, we discuss production of boosted dark matter followed by its expected flux near the Earth in the first subsection.
The next subsection focuses on the scattering of the boosted dark matter in the Earth to create a long-lived particle.
Finally, a detailed discussion on the decay of the long-lived particle appears in the last subsection.

\subsection{Benchmark model and production of boosted dark matter}

As briefly mentioned in the Introduction, the minimal particle contents include super-heavy dark matter $\phi$, light dark matter $\chi_1$, dark-sector unstable state $\chi_2$, and dark gauge boson $X$.
For simplicity, we assume that $\phi$ is scalar while $\chi_{1,2}$ are Dirac-fermionic.
The interaction Lagrangian $\mathcal{L}_{\rm int}$ contains the following operators:
\bea
\mathcal{L}_{\rm int} &\supset& y_\phi \phi \bar{\chi}_1 \chi_1  \label{eq:decay}\\
&& -\frac{\epsilon}{2}F_{\mu\nu}X^{\mu\nu}+(g_{12}\bar{\chi}_2\gamma^\mu \chi_1 X_\mu + {\rm h.c.}), \label{eq:scat}
\eea
where $X_{\mu\nu}$ ($F_{\mu\nu}$) is the field strength tensor for the dark gauge boson $X$ (the ordinary SM photon) and $\epsilon$ is the usual kinetic mixing parameter.
The term in \eqref{eq:decay} governs the decay of $\phi$ to a $\chi_1$ pair and a Yukawa coupling $y_\phi$ parameterizes the decay strength.\footnote{To ensure sufficient stability of $\phi$, $y_\phi$ should be extraordinarily small. The clockwork mechanism may generate dynamically such a tiny coupling constant~\cite{clock}. }
On the other hand, the second term in \eqref{eq:scat} takes care of the up-scattering process of $\chi_1$ to $\chi_2$ and the decay of $\chi_2$ to $\chi_1$ and $X$, with $g_{12}$ encoding their strength.

We emphasize that in the above model the heavier dark matter species $\phi$ does not have direct couplings to SM particles, whereas the lighter one $\chi_1$ has direct interactions through a dark gauge boson portal.
Therefore, any loop-induced decay modes of $\phi$ to SM particles are suppressed enough for $\phi$ to predominantly decays to a pair of $\chi_1$'s.

Speaking of dark matter relics, we assume that $\phi$ dominates over $\chi_1$, so the dark-matter halo in our galaxy essentially consists of $\phi$.
Be aware that the $\phi$ dominance is, in principle, not a necessary condition to explain the ANITA events. The reduction of the $\phi$ fraction would decrease the flux of $\chi_1$ linearly, but it can be compensated by lessening the lifetime of $\phi$ (as shown in Eq.~\eqref{eq:flux}) as far as such a trade-off is consistent with various astrophysical and cosmological observations.
We further imagine the situations where the mass of $\chi_1$, $m_1$ is of MeV to GeV-range, so it can be thermally produced in the early universe.
The present-day relic density of such $\chi_1$ is determined by interrelationships among $\epsilon$, $m_1$, and the mass of dark gauge boson $m_X$.
The relic of $\chi_1$ can be subdominant in a wide range of parameter space as long as $m_1 > m_X$ (and in the resonance regime $2 m_1 \simeq m_X$), and therefore our aforementioned assumption is readily satisfied.
On the other hand, EeV-scale super-heavy $\phi$ under consideration is  produced non-thermally~\cite{Chung:1998zb,Chung:1998ua}, and it is possible for such $\phi$ to take care of almost entire relic abundance of cold dark matter.

Since relic dark matter is mostly given by $\phi$, it is reasonable to assume that the spatial distribution of $\phi$ obeys standard dark matter halo density distributions.
Our choice in this study is the Navarro-Frenk-White profile~\cite{Navarro:1995iw,Navarro:1996gj}, so the $\phi$ energy density $\rho_\phi$ has the form of
\bea
\rho_\phi(r)=\rho_0 \frac{(r/r_s)^{-1}}{(1+r/r_s)^2}\,,
\eea
where $\rho_0\simeq 0.3$ GeV/cm$^3$ is the local dark-matter density at $r\simeq 8.33$ kpc and where $r_s=24$ kpc is the scale radius.
The flux of boosted dark matter $\chi_1$, $\mathcal{F}_1$, for a given direction is then expressed as follows:
\bea
\mathcal{F}_1(\theta_d, \phi_{ra})=\frac{2}{4\pi} \int_{\rm los}ds \frac{1}{\tau_\phi}\cdot \left( \frac{\rho_\phi[r(s,\theta_d,\phi_{ra})]}{m_\phi} \right),  \label{eq:flux}
\eea
where $\tau_\phi$ (or equivalently the inverse of decay rate $\Gamma_\phi$) is the mean lifetime of $\phi$ and where the prefactor 2 takes into account the fact that $\phi$ decays to a pair of $\chi_1$'s.
Here the incident direction is parameterized by the declination angle $\theta_d$ and the right-ascension angle $\phi_{ra}$ in the International Celestial Reference System (ICRS).
We remark that the radial coordinate $r$ in $\rho_\phi$ is a function of these angles and the line-of-sight $s$, and so is the dark-matter energy density $\rho_\phi$.

\subsection{Scattering-off of boosted dark matter}

Once such a boosted $\chi_1$ reaches the Earth, it may scatter off a nucleon via a $t$-channel exchange of dark gauge boson $X$ and a dark-sector unstable state $\chi_2$ comes out.
The nucleon usually breaks up due to a large energy transfer, so the process is essentially initiated with deep inelastic scattering (DIS) between $\chi_1$ and nucleon $N$.
The DIS of this sort is extensively explored in Ref.~\cite{evsp}, and we simply quote their final result for $\sigma_{\chi_1N}^{\rm DIS}$:
\begin{widetext}
\begin{equation}
\frac{d^2\sigma_{\chi_1N}^{\rm DIS}}{dx\,dy} = \frac{\alpha \epsilon^2 g_{12}^2  f(x)}{2 E_1 (Q^2+m_{X}^2)^2} \{  2E_1^2 m_N x (2-2y+y^2)  - m_N x (m_1-m_2)^2 +\left[(m_1^2-m_2^2)(2-y)-2m_1 m_2 y -2m_N^2 x^2 y\right]E_1  \},
\end{equation}
\end{widetext}
where $m_N$ denotes the nucleon mass of interest, $\alpha$ is the usual fine structure constant, and $f(x)$ is the associated parton distribution function (PDF).
The relations among a dimensionful quantity $Q^2$ and two dimensionless quantities $x$ and $y$ are given by
\bea
Q^2&=& -(p_1-p_2)^2\,, \\
y &=& 1-\frac{E_2}{E_1}\,, \\
x &=& \frac{Q^2}{2m_N E_1 y}\,,
\eea
where $p_1$ and $p_2$ denote four momenta of the incoming $\chi_1$ and the outgoing $\chi_2$, respectively.
$x$ implies the momentum fraction carried by a parton while $y$ is fractional energy loss.

For a given $E_1$ one can calculate numerically the $\chi_1$-nucleon DIS cross sections, convolving relevant PDFs.
In our analysis we model the total DIS cross section for a convenient parameter survey, after calculating $\sigma_{\chi_1N}^{\rm DIS}$ of several representative parameter points with \texttt{MSTW2008NNLO}~\cite{Martin:2009iq}.
We find that the following model describes functional behaviors of $\sigma_{\chi_1N}^{\rm DIS}$ in $0.1~{\rm EeV} \lesssim E_1 \lesssim 10~{\rm EeV}$ and $m_X \lesssim 10~{\rm GeV}$ fairly well:
\bea \label{eq:crosssection}
\sigma_{\chi_1N}^{\rm DIS} &\approx& 2.89\times 10^{-29} {\rm cm}^2 \nonumber \\
&\times& \epsilon^2 g_{12}^2 10^{K(\log_{10}[m_X/{\rm GeV}])}\cdot \left( \frac{E_1}{\rm 5~EeV}\right)^{1.05}\,, \nonumber \\
\eea
where $K(x)$ has the form of
\begin{equation}
K(x)=-0.349 x^3-0.452 x^2 - 0.205 x\,.
\end{equation}
Note that the above empirical model does not depend on $m_1$ and $m_2$. This is because our numerical study suggests that the variation in $\sigma_{\chi_1N}^{\rm DIS}$ be of order (at most) $2-3\%$ with $m_{1,2} \lesssim 5~{\rm GeV}$.
As we shall show in Sec.~\ref{sec:results}, some parameter choices belonging to these ranges look highly plausible in generating ANITA-like events.
So, we slenderize our parameter survey, relying on the above-given model and focusing on the associated parameter regions.

Given the cross section, one can estimate the mean free path of $\chi_1$. Kinetic theory suggests that the average travel distance of $\chi_1$ (henceforth denoted by $L_1$ should be
\beq
L_1 \sim \frac{1}{\langle n(\theta_{\rm em})\rangle \cdot \sigma_{\chi_1N}^{\rm DIS}}\,, \label{eq:mfp}
\eeq
where $\langle n \rangle$ is the mean nucleon number density along the $\chi_1$ propagation chord.
Since it depends on Earth layers that $\chi_1$ traverses, we explicitly express its dependence on the emergence angle $\theta_{\rm em}$.
Certainly, the functional behavior of $\langle n \rangle$ in $\theta_{\rm em}$ is highly nontrivial, as the associated density profile inside the Earth shows a drastic change especially at the boundary between adjacent interior layers.
Just to build up our intuition, we find that
\bea
L_1 &\approx& 143,000~{\rm km}\left(\frac{10^{-3}}{\epsilon} \right)^2 \left( \frac{1}{g_{12}}\right)^2 \nonumber \\
&\times& 10^{-K(\log_{10}[m_X/{\rm GeV}])}\cdot \left( \frac{\rm 5~EeV}{E_1}\right)^{1.05}
\eea
for $\theta_{\rm em}=30^\circ$ at which the chord length is the same as the Earth radius.

\subsection{Decay of a long-lived particle}

Once $\chi_2$ is created, it decays back to $\chi_1$ together with hadrons (through a quark pair) which invoke EAS.
As mentioned in the Introduction, the $\chi_2$ decay takes place via either on-shell or off-shell dark gauge boson, depending on the mass relation among $\chi_1$, $\chi_2$ and $X$.
Labeling them by ``on-shell'' scenario and ``off-shell'' scenario respectively, we discuss their characteristic features separately.

\medskip

\noindent {\bf On-shell scenario}: In this scenario $m_2$ is greater than $m_1+m_X$ so that an on-shell $X$ comes out as a decay product of $\chi_2$.
Unless the coupling $g_{12}$ and/or a mass gap $m_2 - m_1-m_X$ is unusually small, the $\chi_2$ decay is prompt (i.e., it occurs at the scattering point), and $X$ may be long-lived and propagating.
The total decay width of $X$, $\Gamma_X$ is given by
\begin{equation}
\Gamma_X =\frac{\epsilon^2 \alpha m_X}{3}\sum_i C_i \left(1+\frac{m_i^2}{m_X^2}\right)\sqrt{1-\frac{4m_i^2}{m_X^2}}\,,
\end{equation}
where $C_i$ denotes the color factor for particle species $i$ which runs over all charged SM fermions lighter than half the mass of $X$.\footnote{A special care must be taken for quarks. Since hadrons appear as decay products, their masses, not quark ones, should be considered accordingly.}
As a rough estimate we see that the laboratory-frame mean decay length of $X$, $\ell_{X,{\rm lab}}$ is
\begin{equation}
\ell_{X,{\rm lab}} \sim 400~{\rm km}\left( \frac{10^{-5}}{\epsilon}\right)^2 \left( \frac{500~{\rm MeV}}{m_X} \right)^2 \left( \frac{E_X}{\rm EeV}\right)\,, \label{eq:ellX}
\end{equation}
with $E_X$ being the amount of energy that $X$ carries.

One possible issue with this scenario is disappearance of $X$ via $X+N \to \hbox{anything}$, before it decays.
To develop some intuition on such a possibility, its cross section should be first estimated.
If $X$ is energetic enough to be treated nearly massless, we can get some hint from the photon-nucleon DIS cross section.
From the experimental measurements of $\sigma_{\gamma p}^{\rm DIS}$ at HERA~\cite{Aid:1995bz,Abramowicz:2010asa} we find
\begin{equation}
\sigma_{X p}^{\rm DIS} \approx 1.65 \times 10^{-36}~{\rm cm}^2\left( \frac{\epsilon}{10^{-4}}\right)^2 \left( \frac{E_{\rm c.o.m.}}{200~{\rm GeV}}\right)^{0.22}\,,
\end{equation}
where $E_{\rm c.o.m.}$, the center-of-mass energy of the $X$--$p$ system, is given by $\sqrt{2E_X m_p}$.\footnote{In Ref.~\cite{Abramowicz:2010asa} the exponent upon $E_{\rm c.o.m.}$ is extracted with data for $E_{\rm c.o.m.}\sim 200-300$ GeV. We assume that this functional behavior does not alter much at $(2\cdot 10^9~{\rm GeV}\cdot 1~{\rm GeV})^{1/2}\approx 45~{\rm TeV}$. }
One may estimate the typical distance for a single scattering process using the relation in~\eqref{eq:mfp}, and see that the distance is much larger than the mean decay length above for a given set of $E_X$ and $\epsilon$.
In other words, dark gauge boson is strongly inclined to decay before disappearing.

\medskip

\noindent {\bf Off-shell scenario}: Contrary to the previous scenario, the mass of dark gauge boson $X$ is larger than the mass difference between $\chi_2$ and $\chi_1$.
Therefore, if the mass gap $\delta m~(\equiv m_2-m_1)$ (kinematically) allows for opening decay modes to SM fermion pairs, the associated decay process occurs via an off-shell $X$.
$\chi_2$ is now long-lived and the associated decay width $\Gamma_2$ is formulated~\cite{Giudice:2017zke} by the form of
\begin{equation}
\Gamma_2 \approx \frac{\epsilon^2\alpha g_{12}^2}{15\pi^2m_X^4}(\delta m)^5 \sum_i C_i\,,
\end{equation}
under the assumption of $m_i \ll \delta m \ll m_2 \ll m_X$ (see Appendix B in Ref.~\cite{Giudice:2017zke} for the exact formula).
Our rough estimate shows that the laboratory-frame mean decay length of $\chi_2$, $\ell_{2,{\rm lab}}$ is
\bea
\ell_{2,{\rm lab}} &\sim& 12,500~{\rm km} \left( \frac{10^{-4}}{\epsilon} \right)^2 \left( \frac{1}{g_{12}} \right)^2 \label{eq:ell2} \\
&\times& \left(\frac{m_X}{\rm 2~GeV} \right)^4 \left( \frac{0.5~{\rm GeV}}{\delta m} \right)^5 \left(\frac{{\rm GeV}}{m_2} \right)\left( \frac{E_2}{\rm EeV} \right), \nonumber
\eea
where $E_2$ is the energy of the scattered $\chi_2$.

\medskip

We close this section, rendering a few comments.
It is summarized that the event topology under consideration begins with an upscattering of an invisible particle and accompanies a long-lived intermediary state decaying to hadrons.
Indeed, the structure of upscatter-decay has been adopted in a diverse range of astrophysical phenomena~\cite{Finkbeiner:2007kk, ArkaniHamed:2008qn, Pospelov:2008jd, Finkbeiner:2009mi, Kim:2017qaw}, search strategies in particle accelerator experiments~\cite{Essig:2007az, ArkaniHamed:2008qp, Bai:2011jg, Bell:2013wua, Izaguirre:2014dua, Izaguirre:2017bqb, Berlin:2018pwi, Berlin:2018jbm}, double-bang events at IceCube~\cite{Coloma:2017ppo, Coloma:2019qqj} and inelastic boosted dark matter~\cite{Kim:2016zjx,Giudice:2017zke,Chatterjee:2018mej,evsp}.
In particular, Refs.~\cite{Pospelov:2013nea, Grossman:2017qzw, Eby:2019mgs} have considered the up-scattering of dark matter to an excited state somewhere outside a detector, followed by its de-excitation in the detector along with visible particles.
In the context of the ANITA anomaly, Refs.~\cite{Collins:2018jpg, Cline:2019snp} have taken related strategies.

\section{Analysis Procedure \label{sec:analysis}}

\begin{figure}[t]
\centering
\includegraphics[width=5cm]{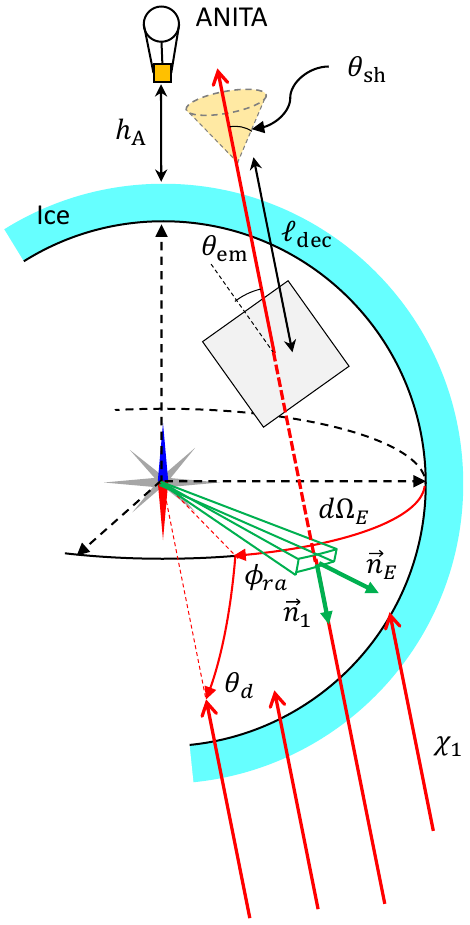}
\caption{\label{fig:coord} The notation adopted in our simulation study.}
\end{figure}

In this section, we detail the procedure that we used in order to compute the differential number of events per emergence angle at which
the ANITA detector would observe events according to our model prediction, redefining notation and conventions more carefully if needed.
Generically, the expected number of signal events $N_{\rm sig}$ is given by the exposure time $t_{\rm exp}$ times the inner product of flux and effective area vectors.
From Eq.~\eqref{eq:flux} we see that our signal flux varies with the angles $\theta_d$ and $\phi_{ra}$ (see also FIG.~\ref{fig:coord} for the notations).\footnote{Note, however, that the flux producing upward-going showers at the south pole corresponds to the decay of DM particle decaying in the outskirt of the galaxy. Therefore the angular distribution of the incoming flux is essentially isotropic in our case. The impact of the choice of DM profile is therefore negligible in our work.}
Labeling the ``effective'' area projected onto the flux vector by $\mathcal{A}_{\rm eff}$, we remark that $\mathcal{A}_{\rm eff}$ potentially depends on $\theta_d$ and $\phi_{ra}$.
Therefore, $N_{\rm sig}$ is obtained after integrating over the incoming flux direction $\theta_d$ and $\phi_{ra}$:
\begin{equation}
N_{\rm sig}= t_{\rm exp}\times \int d\Omega_{\rm ICRS}~ \mathcal{A}_{\rm eff}(\theta_d,\phi_{ra})~\mathcal{F}_1(\theta_d,\phi_{ra})\,,
\end{equation}
where $d\Omega_{\rm ICRS}=d(\cos\theta_d)d\phi_{ra}$.
Note that in our model the energy of the incoming boosted dark-matter state, $E_1$, is uniquely determined by the mass of the heavy dark-matter state $\phi$.
Therefore, the flux of $\chi_1$ particles, $\mathcal{F}_1$, is implicitly depending on $E_1$.
Likewise, (as we will see later) $\mathcal{A}_{\rm eff}$ varies with $E_1$, so $N_{\rm sig}$ is essentially a function of the energy $E_1$. For brevity, we will omit the $E_1$ dependence in what follows, unless it is necessary for a better understanding.
We further note that for a given energy $E_1$ the effective area differs according to the emergence angle $\theta_{\rm em}$ with which the decaying particle (either $X$ or $\chi_2$) escapes the surface of the Earth and according to its energy $\mathcal{E}_{\rm ex}$ when it exits the Earth crust.
Therefore, for a given direction of the flux arrival $(\theta_d,\phi_{ra})$, the effective area $\mathcal{A}_{\rm eff}(\theta_d,\phi_{ra})$ can be formulated as
\begin{equation}
\mathcal{A}_{\rm eff}(\theta_d,\phi_{ra})=\int d\theta_{\rm em} d\mathcal{E}_{\rm ex} A_{\rm eff}(\theta_d,\phi_{ra};\theta_{\rm em},\mathcal{E}_{\rm ex})\,,
\end{equation}
where
\begin{equation}
A_{\rm eff}(\theta_d,\phi_{ra};\theta_{\rm em},\mathcal{E}_{\rm ex})\equiv \frac{d^2 \mathcal{A}_{\rm eff}(\theta_d,\phi_{ra})}{d\theta_{\rm em} d\mathcal{E}_{\rm ex}}\,. \label{eq:Aeff}
\end{equation}
Since one of the main goals in our work is to obtain an appropriate angular distribution of the signal as seen by ANITA, we seek to obtain the differential number of signal events in $\theta_{\rm em}$, i.e.,
\begin{equation}\label{eq:dNdtheta}
\hspace{-0.2cm}\frac{dN_{\rm sig}}{d\theta_{\rm em}}=t_{\rm exp}\int d\Omega_{\rm ICRS}d\mathcal{E}_{\rm ex} A_{\rm eff}(\theta_d,\phi_{ra};\theta_{\rm em},\mathcal{E}_{\rm ex}) \mathcal{F}_1(\theta_d,\phi_{ra}).
\end{equation}

The key quantity in order to compute this angular distribution is the elementary effective area $A_{\rm eff}$ in Eq.~\eqref{eq:Aeff} per emergence angle $\theta_{\rm em}$ and exit energy $\mathcal{E}_{\rm ex}$ for a given direction $d\vec{\Omega}_{\rm ICRS}$ and energy $E_1$ of the incoming flux of $\chi_1$ particles.
This area essentially encodes all the dynamical details of the detection, from the $\chi_1$ entrance in the Earth, its propagation and conversion into a long-lived state, to the possible decay of the latter in the atmosphere and the EAS detection by the radio-antennas of ANITA.
Inspired by the analysis scheme in Ref.~\cite{Romero-Wolf:2018zxt}, we modulize the whole calculational procedure in the following fashion: for a given incoming energy and direction of the flux, we
\begin{itemize}
\item [$(i)$] integrate over the entire Earth surface, scanning over all the surface points of impact where a flux of $\chi_1$ particles arriving along parallel chords strikes;
\item [$(ii)$] compute the probability that the $\chi_1$ propagates through the Earth, scatters and creates a long-lived particle (either $X$ or $\chi_2$) which escapes the Earth surface with a given energy $\mathcal E_{\rm ex}$;
\item [$(iii)$] compute the probability that the long-lived particle decays in the atmosphere at a relevant height, creating an EAS;
\item [$(iv)$] calculate the probability that such shower is detected by ANITA.
\end{itemize}
Such decomposition can be formulated as
\begin{eqnarray}
&A_{\rm eff}& \hspace{-0.15cm}(\theta_d,\phi_{ra};\theta_{\rm em},\mathcal{E}_{\rm ex})= \nonumber \\ [1em]
&&
\begin{array}{l l}
 R_E^2\int d\Omega_E \vec{n}_E \cdot \vec{n}_1 & \hbox{ for $(i)$} \\ [1em]
\times \frac{d}{d\mathcal{E}_{\rm ex}} P_{\rm ex}(E_1;\theta_d,\phi_{ra};\theta_{\rm em}, \theta_E,\phi_E) & \hbox{ for $(ii)$} \\ [1em]
\times \int d\ell_{\rm dec} \frac{dP_{\rm dec}(\mathcal{E}_{\rm ex})}{d\ell_{\rm dec}} &\hbox{ for $(iii)$}  \\ [1em]
\times P_{\rm det}(\mathcal{E}_{\rm EAS},\theta_{\rm sh},\ell_{\rm dec};\theta_d,\phi_{ra};\theta_E,\phi_E), &\hbox{ for $(iv)$}
\end{array} \nonumber \\
&& \label{eq:Aefffull}
\end{eqnarray}
where $\vec{n}_E$ is the Earth-surface unit normal vector while $\vec{n}_1$ is a unit vector lying in the flux direction of $\chi_1$ (see also FIG.~\ref{fig:coord}).
The solid angle element with respect to the Earth center is denoted by $d\Omega_E = d(\cos\theta_E)d\phi_E$.
In the third row we explicitly express the dependence of the differential exit probability $dP_{\rm ex}/d\mathcal{E}_{\rm ex}$ on $E_1$ as well.
In the fourth row the decay probability of the long-lived particle is integrated along its propagation chord.
Finally, in the last row $P_{\rm det}$ describes the capability for a given EAS to yield a sufficient electric field along a direction which would reach the payload hence result in detection. We mark its dependence on the shower opening angle $\theta_{\rm sh}$ and the energy of EAS $\mathcal{E}_{\rm EAS}$.
The ensuing subsections are devoted to discussing in more details the calculation of the steps $(ii)$ through $(iv)$.

\subsection{Exit probability}

Throughout this work, we perform a Monte Carlo simulation in order to estimate accurately the exit probability $P_\text{ex}$ appearing in the third row of Eq.~\eqref{eq:Aefffull}, as a function of the emergence angle $\theta_\text{em}$.
To this end, we adapted the public code provided with Ref.~\cite{Alvarez-Muniz:2017mpk} for our BSM scenario.
Assuming the Earth surface to be covered by a 4~km-thick ice layer, the simulation accounts for the varying density of the Earth crust along the propagation chord.

The quantity $dP_{\rm ex}(E_1;\theta_d,\phi_{ra};\theta_{\rm em}, \theta_E,\phi_E)/d\mathcal{E}_{\rm ex}$ is defined as the probability that an incoming boosted dark-matter particle $\chi_1$ -- which arrives with an incidence angle $\theta_i$ (corresponding to an emergence angle $\theta_\text{em}=90^\circ-\theta_i$) and an energy $E_1$ -- exits the Earth with an energy $\mathcal E_\text{ex}$.
For every emergence angle $\theta_\text{em}$, each event of the Monte Carlo simulation therefore tracks along the propagation chord the different possible interactions, according to the two aforementioned benchmark scenarios.

\medskip

\noindent{\textbf{On-shell scenario:}}
\begin{enumerate}
\item An incoming $\chi_1$ particle scatters off a nucleon with the cross-section appearing in Eq.~\eqref{eq:crosssection} and converts into a particle $\chi_2$ which promptly decays into a dark photon $X$ and a boosted particle $\chi_1$,
\item the dark photon $X$ propagates until it reaches the Earth surface, as long as it does not decay inside the Earth, and
\item the (secondary) boosted dark-matter particle $\chi_1$ produced by the $\chi_2$ decay propagates until it re-scatters inside the Earth or escapes the surface.
\end{enumerate}

\medskip

\noindent{\textbf{Off-shell scenario:}}
\begin{enumerate}
\item An incoming $\chi_1$ particle scatters off a nucleon with the cross-section shown in Eq.~\eqref{eq:crosssection} and converts into particle $\chi_2$,
\item the dark-sector particle $\chi_2$ propagates until it decays inside the Earth or escapes the surface, and
\item in the case that $\chi_2$ decays inside the Earth, its decay product (secondary) $\chi_1$ propagates until it re-scatters inside the Earth or escapes the surface.
\end{enumerate}

Pursuing the propagation of secondary $\chi_1$'s created in the Earth by the $\chi_2$ decay allows us to keep track of possible regeneration which could potentially increase the exit probability at large emergence angles and low exit energies.
However, our simulation study suggests that the regeneration play an insignificant role in the scenarios that we investigate in this work, as we have elaborated in Sec.~\ref{sec:transparency}.
We therefore focus on cases where the Earth is relatively transparent to the flux of incoming particles.

\subsection{Decay probability in the atmosphere}

Once a long-lived particle species $i~(=X \hbox{ or } \chi_2)$ comes out of the Earth surface, its decay simply obeys the exponential decay law
\beq
\frac{dP_{\rm dec}(\mathcal{E}_{\rm ex})}{d\ell_{\rm dec}} = \frac{1}{\ell_{i,\rm lab}}e^{-\ell_{\rm dec} /\ell_{i,\rm lab}}\,,
\eeq
where $\ell_{i,\rm lab}$ is the laboratory-frame mean decay length of the long-lived particle $i$.
In the two cases where $i=X$ and $i=\chi_2$, the results appearing in Eqs.~\eqref{eq:ellX} and \eqref{eq:ell2} can be substituted to $\ell_{i,\rm lab}$ with $E_X$ and $E_2$ replaced by $\mathcal{E}_{\rm ex}$.\footnote{Be aware again that Eq.~\eqref{eq:ell2} is an approximation valid under restricted mass spectra. Our simulation study was based on the exact formula in Ref.~\cite{Giudice:2017zke}.}

As shown in Ref.~\cite{Heurtier:2019git}, in the case where an astrophysical flux would produce tau leptons inside the Earth, the exit probability of the tau leptons would spread over a wide range of exit energies.
Indeed, taus quickly lose energy while propagating in the Earth.
Moreover, during regeneration processes, the particle produced through a decay, which may then re-scatter off a nucleon, will lead to an event with even lower exit energy.
In our case, this sort of energy loss of the particles propagating through the Earth can be negligible since their interactions with the visible sector particles are very feeble.
We therefore take into consideration only the energy losses (with respect to incoming energy $E_1$), which are caused by the fractional energy release in the scattering and decay processes.

The value of the exit energy has a crucial impact on the decay probability since the more boosted particles are, the less they are expected to decay before reaching the ANITA detector.
Our simulation carefully takes care of possible energy diminishment down to $\mathcal{E}_{\rm ex} = 100$ TeV, although most of the events involve a much larger exit energy in our proposed model due to the reasons discussed above.

\subsection{Detection probability}

The hadronic decay products can create an EAS in the atmosphere which can eventually be detected by the radio-antennas of ANITA at an altitude of $h_A\sim 35~\mathrm{km}$.
Experimentally, it is not sufficient that the shower reaches the payload but the local electric field measured by the ANITA detector should be large enough to overcome the associated threshold.
Here the probability $P_{\rm det}$ parameterizes the likelihood of recording an event with such effects considered.

The authors of Ref.~\cite{Romero-Wolf:2018zxt} have performed a dedicated analysis regarding this probability, drawing a few important observations: $(a)$ it gets gradually challenging for a shower created at altitudes above $\mathcal{O}(10~{\rm km})$ to induce a fully-developed EAS, $(b)$ the electric field peak, depending on the altitude where the shower is produced, lies at an opening angle $\theta_{\rm sh}$ in-between $\sim 1^\circ$ and $\sim 2^\circ$, $(c)$ the lower $\mathcal{E}_{\rm ex}$ is, the less the ANITA detector is triggered, and $(d)$ the further from the ANITA detector the EAS is produced, the less likely it is to be detected.

Unfortunately, although Ref.~\cite{Romero-Wolf:2018zxt} provides an approximate formula in order to evaluate the electric field produced by an EAS at the ANITA location, with an emergence angle of $30^\circ$ and arbitrary distance from the point of decay and shower angle, the exact functional dependence of the parameters involved in this expression  with the emergence angle has not been exhibited. This renders the evaluation of the precise probability $P_{\rm det}$ challenging from a theoretical point of view.
For simplicity, we rather use the main conclusions of Ref.~\cite{Romero-Wolf:2018zxt}, following the assumptions made in Ref.~\cite{Heurtier:2019git}, that is, we assume that $P_{\rm det}=1$ as long as the shower is produced at an altitude lower than 15~km, and whenever the detector is contained in a cone of angular opening $\theta_{\rm sh} <1.5^\circ$ with respect to the shower axis, and $P_{\rm det}=0$ otherwise.

Finally, the $\mathcal{E}_{\rm EAS}$ dependence of $P_{\rm det}$ must be taken with care, as the energy of the EAS could potentially be related to $\mathcal{E}_{\rm ex}$ in a non-trivial manner. In the ``on-shell'' scenario, the whole energy of the escaping $X$ is transferred to decay products as it decays fully visibly. Therefore, we are allowed to set $\mathcal{E}_{\rm EAS}=\mathcal{E}_{\rm ex}$.
On the contrary, in the ``off-shell'' scenario, a certain fraction of $\mathcal{E}_{\rm ex}$ is taken away by the outgoing $\chi_1$.
We find that if $\chi_{1,2}$ are sufficiently heavier than hadronic decay products, $\chi_1$ is produced nearly at rest in the $\chi_2$ rest frame. Therefore, in the laboratory frame, $\chi_1$ takes away $\sim (m_1/m_2) \mathcal{E}_{\rm ex}$, and our simulation study with \texttt{PYTHIA 8}~\cite{Sjostrand:2014zea} confirms that a large majority of events conforms to this estimate.
In our analysis with off-shell scenario, we survey parameter space satisfying such mass relations, and thus adopt the approximation that $\mathcal{E}_{\rm EAS} \approx (\delta m/m_2)\mathcal{E}_{\rm ex}$.

\section{Results and Interpretations \label{sec:results}}

\begin{figure*}[t]
\begin{center}
\includegraphics[width=\linewidth]{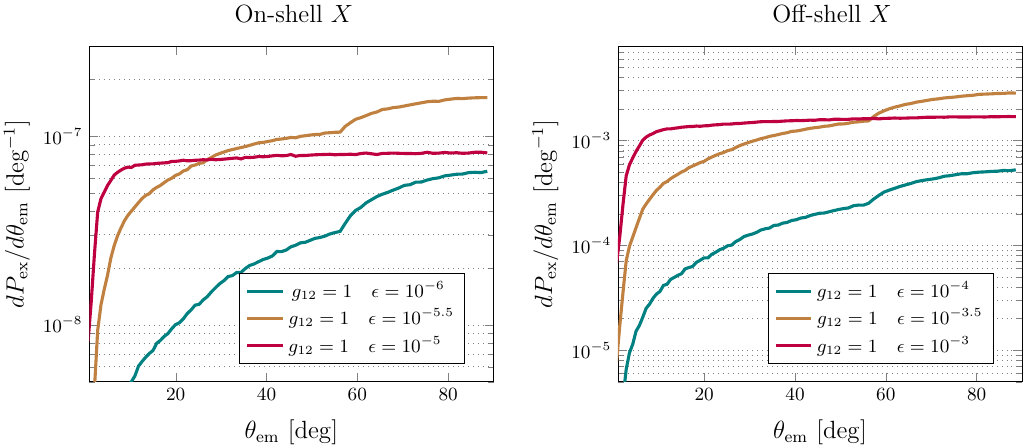}
\caption{\label{fig:MonteCarlo} $\mathcal{E}_{\rm ex}$-integrated differential exit probability in emergence angle $\theta_{\rm em}$ for the ``on-shell'' scenario (left panel) an the ``off-shell'' scenario (right panel).
Mass parameter choices are summarized in~\eqref{eq:masschoice}, while the values of coupling constants $\epsilon$ and $g_{12}$ are displayed in the legends. }
\end{center}
\end{figure*}

We now report our simulation results in this section, followed by discussing their phenomenological implications.
In order to demonstrate that the inelastic boosted dark matter scenario can accommodate the ANITA anomalous events,
we choose the following two benchmark sets of parameters for the on-shell and off-shell cases.
\begin{eqnarray}
\textbf{On-shell}\text{: }&& m_{2}>m_{1}+m_X,~m_X=0.5~\mathrm{GeV},\nonumber\\
&&  m_{\phi}= 2~{\rm EeV}, \nonumber \\
\textbf{Off-shell}\text{: }&&  m_{2}=2.5~\mathrm{GeV},~m_{1}=2~\mathrm{GeV},~m_X=2~\mathrm{GeV},\nonumber \\
&&  m_{\phi}= 4~{\rm EeV}. \label{eq:masschoice}
\end{eqnarray}
Note that in the on-shell case the masses of $\chi_1$ and $\chi_2$ do not play any role either in the simulation of the effective area or in the decay width of the dark gauge boson, as long as they satisfy the mass relation with $X$, they are light enough to be considered relativistic, and $\chi_2$ decays instantly.
By contrast, in the off-shell scenario, all the mass values affect the propagation and the decay, although $m_1$ does not play a role in the propagation code.
The $m_\phi$ values are selected in such a way that the associated hadronic decay products collectively carry away an energy within $0.5-1$ EeV.

\begin{figure*}[t]
\centering
\includegraphics[width=0.9\linewidth]{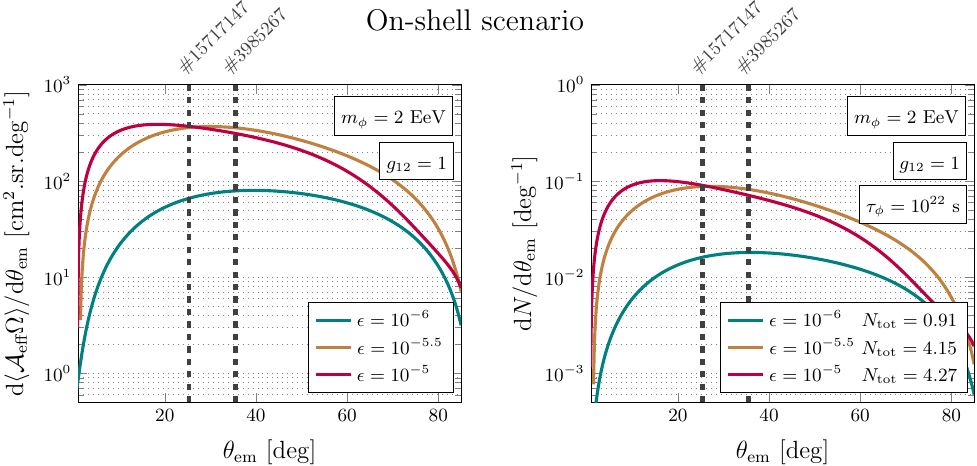}

\vspace{0.5cm}
\includegraphics[width=0.9\linewidth]{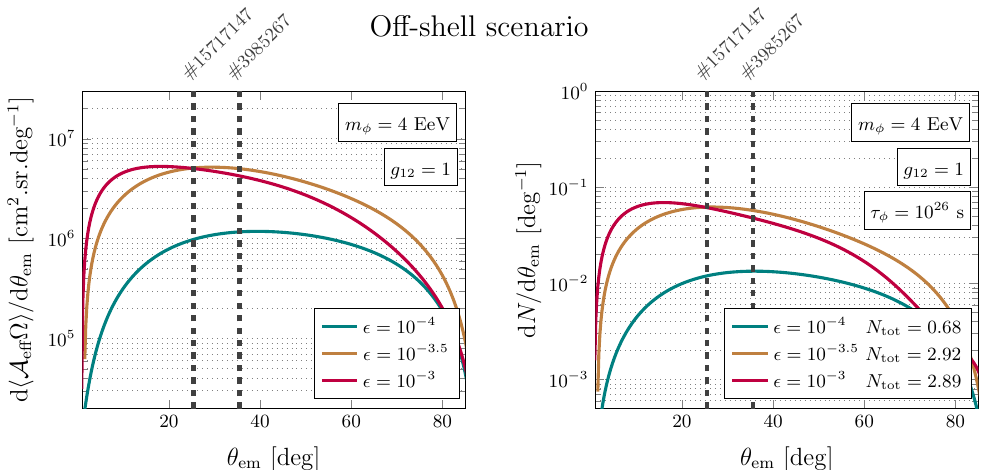}
\caption{\label{fig:dNdtheta} Total effective area (left panels) and differential number of events (right panels) as a function of the emergence angle $\theta_{\rm em}$ anticipated for ANITA-85.5 days, in the cases of ``on-shell'' scenario (upper panels) and ``off-shell'' scenario (lower panels).
Parameter choices appear in the respective legends. }
\end{figure*}

In FIG.~\ref{fig:MonteCarlo} we exhibit the differential exit probability in the emergence angle $\theta_{\rm em}$, integrating over all possible exit energy $\mathcal{E}_{\rm ex}$.
The left and right panels correspond to the on-shell and off-shell scenarios, respectively, for which we vary the value of the kinetic mixing parameter $\epsilon$ with the value of the dark-sector coupling $g_{12}$ fixed to unity.
We see that the red curves in both scenarios show a different behavior from the others.
As we elaborated in Sec.~\ref{sec:transparency}, this is because the mean decay length of $X$ (for the on-shell scenario in the left panel) or $\chi_2$ (for the off-shell scenario in the right panel) resulting from the associated parameter choices is much shorter than most of the chord lengths so that the exit probability quickly saturates in increasing $\theta_{\rm em}$.

We remark that the variation in $g_{12}$ would alter the results in FIG.~\ref{fig:MonteCarlo} in different manners in the on-shell and off-shell scenarios.
In the former case, raising $g_{12}$ would result in enhancing the scattering cross section, while the decay length of $X$ remains unaffected.
As far as the mean free path of $\chi_1$ keeps much larger than the dimension of the Earth, this would simply lead the effect of rescaling the total exit probability by $g_{12}^2$.
On the contrary, in the latter case, both the scattering cross section and the decay width of $\chi_2$ scale as $g_{12}^2\epsilon^2$.
Therefore, increasing or decreasing the dark-sector coupling is equivalent to inversely varying the kinetic mixing parameter by the same magnitude.

Following the procedure described in Sec.~\ref{sec:analysis}, we next  integrate over the incoming $\chi_1$ directions defined by the $(\theta_d,\phi_{ra})$ pair and the Earth surface $d\Omega_E$, in order to find the effective area of the detector.
The resulting plots for the on-shell and off-shell scenarios are demonstrated in the upper-left panel and the lower-left panel of FIG.~\ref{fig:dNdtheta}, respectively.
We then convolve the results with the $\chi_1$ flux expected from the decay of the heavy dark-matter particle $\phi$ [see also Eq.~\eqref{eq:flux}] and total exposure time of the three flights of ANITA, in order to calculate the differential number of events per emergence angle, i.e., Eq.~\eqref{eq:dNdtheta}.
The associated plots for the on-shell and off-shell scenarios appear in the upper-right panel and the lower-right panel of FIG.~\ref{fig:dNdtheta}, correspondingly.
For reference purpose, we mark the $\theta_{\rm em}$ values associated with the two ANITA events by black dashed vertical lines.

In both scenarios we adapt the lifetimes of the decaying particles such that the total expected number of events is of $\mathcal O(1)$, given the exposure time $t_{\rm exp}=85.5~\mathrm{days}$.
As can be seen in both the on-shell and off-shell cases, varying the value of the kinetic mixing parameter clearly exhibits the transition from the opaque case (large $\epsilon$) to the transparent case (small $\epsilon$).
Furthermore, we observe that the maximum value of the angular distribution changes according to what was described in Sec.~\ref{sec:transparency}.
As expected previously, the intermediate case (or translucent case) shows a maximum at emergence angles of $\mathcal O(25^\circ-35^\circ)$, which is the range in which the two anomalous events observed by ANITA are lying.

Finally, we make a brief comment on the relic abundance of $\chi_1$.
Since our benchmark dark-sector model assumes that $\chi_1$ is a negligible relic component, it is crucial to check whether or not our parameter choices indeed advocate the assumption.
We have explicitly calculated the $\chi_1$ relic density for all sets of parameter choices, employing \texttt{MicrOMEGAs}~\cite{Belanger:2018mqt}.
The numerical outcomes suggest that $m_1 \gsim 0.45$ GeV in the on-shell scenario and $m_1 \gsim 1.9$ GeV in the off-shell scenario be good enough to have negligible ($\lsim 0.1\%$) abundance for our coupling choices, in addition to the resonance regime satisfying $m_X \simeq 2m_1$.

\section{Conclusions and Outlook \label{sec:conclusion}}

The upward-moving anomalous events reported recently by the ANITA Collaboration~\cite{Gorham:2016zah,Gorham:2018ydl} have brought increasing attention in the particle physics community, as it seems disfavored to explain them under the SM framework, e.g., tau-neutrino-induced processes.
Therefore, the phenomenon has been considered as a hint to new physics beyond the SM, stimulating a growing number of phenomenological studies in the context of new physics scenarios~\cite{Fox:2018syq,Cherry:2018rxj,Huang:2018als,Anchordoqui:2018ucj,Chauhan:2018lnq,Anchordoqui:2018ssd,Collins:2018jpg,Heurtier:2019git,Cline:2019snp,Hooper:2019ytr}.
In this paper, we conducted a dedicated study on interpreting the ANITA anomaly with scenarios of inelastic boosted dark matter~\cite{Kim:2016zjx}.

The proposed model in this paper includes an EeV-range super-heavy (scalar) dark matter $\phi$ which predominantly decays to a pair of MeV-range (fermionic) dark matter particles $\chi_1$ in the galactic halo in the present universe.
Due to the large mass difference between the two dark-matter states, the lighter dark-matter particles $\chi_1$ produced by the decay of the super-heavy candidate $\phi$ are highly boosted and reach the Earth.
An incident $\chi_1$ may therefore upscatter to a heavier, unstable dark-sector state $\chi_2$ via an exchange of dark gauge boson $X$. $\chi_2$ then decays to $\chi_1$ and hadronic final states through on-shell (off-shell) intermediary $X$, while $X$ ($\chi_2$) often becomes considerably long-lived.
Such a long-lived particle may escape from the Earth, carrying an EeV-scale energy.
Once it decays in the atmosphere at a proper altitude, there arises an EAS which can be detected by ANITA.

We then conducted a parameter survey according to our analysis strategy elaborated in Sec.~\ref{sec:analysis}.
We examined opaque, translucent, and transparent cases (see Sec.~\ref{sec:transparency} for their definitions) with different parameter choices, and found that the translucent case can accommodate the ANITA events best in the sense that the expected number of signal events is maximized at an emergence angle in-between $25^\circ$ and $35^\circ$.
The selected parameter values are consistent with existing constraints.
More importantly, the dark-matter relic abundance associated with our parameter choices supports the benchmark dark-sector model assumption that $\phi$ and $\chi_1$ are the dominant and negligible relic components, respectively.
Therefore, the anomalous events observed by the ANITA Collaboration would be considered as a sign of the inelastic boosted dark matter scenario adopted in this study.

We would like to emphasize that in our scenario, the very feeble interaction of the particles propagating through the Earth with nuclei renders detectors such as the Auger observatory or IceCube far less competitive than ANITA in terms of signal detection, given their small fiducial volume as compared to the volume of the South Pole atmosphere that ANITA utilizes to observe EAS.
Indeed, as opposed to scenarios which require to produce boosted tau leptons inside the Earth crust in order to produce EAS in the atmosphere, our scenario simply requires that boosted $X$ or $\chi_2$ particles reach the Earth surface before decaying, hence very unlikely to leave any significant traces in Earth-based detector.

As a final remark, we point out that although our approach is predicated upon a non-minimal boosted dark matter model containing a dark-sector unstable state ($\chi_2$ in this study), scenarios similar to our ``on-shell'' scenario are available in the minimal boosted dark matter model, by allowing for the following term in the Lagrangian:
\beq
\mathcal{L}_{\rm int} \supset g_{11}\bar{\chi}_1 \gamma^\mu \chi_1 X_\mu\,,
\eeq
where $g_{11}$ is the coupling constant governing a $\chi_1$-flavor-conserving interaction.
The existence of this operator, in turn, allows for the process
\beq
\chi_1 + N \to \chi_1 + X + {\rm anything}\,,
\eeq
by radiating a dark gauge boson $X$ off initial-/final-state $\chi_1$ (called ``dark-strahlung'')~\cite{Kim:2019had}.
This scenario is plausible and the event rate would be sizable especially for smaller $m_X$ and larger $g_{11}$, compared to those of the current reference parameter sets. 
It is certainly interesting to distinguish the two possibilities which requires a dedicated study. 
We leave the exploration of this direction for a future work.

\section*{Acknowledgments}


We would like to thank Kim Berghaus, Bhupal Dev, Ian Lewis, Yann Mambrini, Mathias Pierre, and Yicong Sui for useful discussions.
The work of L.H. and D.K. is supported by the Department of Energy under Grant DE-FG02-13ER41976 (de-sc0009913).
The work of J.-C.P. is supported by the National Research Foundation of Korea (NRF-2019R1C1C1005073 and NRF-2018R1A4A1025334).
The work of S.S. is supported by the National Research Foundation of Korea (NRF-2017R1D1A1B03032076 and in partial by NRF-2018R1A4A1025334).
A part of this work was discussed in the workshop ``Dark Matter as a Portal to New Physics" at Asia Pacific Center for Theoretical Physics in Jan 14th - 18th, 2019.

\bibliographystyle{apsrev4-1}
\bibliography{draft}

\end{document}